\documentclass[aip,reprint]{revtex4-1}
\usepackage{graphicx} 
\usepackage{color}
\definecolor{mycol}{RGB}{0.5,0.5,0}

\draft 

\begin{document}


\title{Reducing current noise in cryogenic experiments by vacuum-insulated cables} 


\author{E. Mykk\"anen}
\author{J. S. Lehtinen}
\author{A. Kemppinen}
\affiliation{VTT Technical Research Centre of Finland Ltd, Centre for Metrology MIKES, PO Box 1000, VTT 02044, Finland}
\author{C. Krause}
\author{D. Drung}
\affiliation{Physikalisch-Technische Bundesanstalt (PTB), Abbestraße 2-12, 10587 Berlin, Germany}
\author{J. Nissil\"a}
\author{A. J. Manninen}
\affiliation{VTT Technical Research Centre of Finland Ltd, Centre for Metrology MIKES, PO Box 1000, VTT 02044, Finland}

\date{\today}

\begin{abstract}
We measure the current noise of several cryogenic cables in a pulse tube based dilution refrigerator at frequencies
between about 1~mHz and 50~kHz. We show that vibration-induced noise can be efficiently suppressed by using
vacuum-insulated cables between room temperature and the 2nd pulse tube stage. A noise peak below 4 fA at the
1.4~Hz operation frequency of the pulse tube and a white noise density of 0.44~fA$/\sqrt{\mathrm{Hz}}$ in the
millihertz range are obtained.

\end{abstract}

\pacs{}

\maketitle

\section{Introduction}

Increasing popularity of pulse tube refrigerators, which generate vibration induced current noise,\cite{TomaruCryogenics2004,RiabzevCryogenics2009} has made the study of minimizing the vibrations and their effect on measurement signals increasingly important. The current noise levels are relevant for example in sub-nanoampere quantum current standard applications.\cite{GiblinNatCom2012,SteinAPL2015,PekolaRevModPhys2013} In some cases the excess current noise may even cause decoherence in quantum systems.\cite{KalraArxiv2016} Special care has been taken to construct pulse tube operated systems that are vibration free.~\cite{WangCryogenics2010,WangAdvCryo2010} However, since these techniques are not yet very widely spread, the study of low-noise cryogenic cables is also important.

Mechanical vibrations can induce current noise in cables through microphonic, tribo-, and piezoelectric effects.\cite{OngEJoP1987,keithleyBook} The first of these arises from the change of cable capacitance, $I_n=V$ d$C/$d$t$, and is often negligible if the voltage $V$, the cable capacitance $C$ and relevant frequency are small. In both tribo- and piezoelectric effects, the insulator of a cable releases or absorbs charge when the cable is bent, e.g.~by pulse tube vibrations. The former effect is due to friction between the insulator and the conductor, and the latter one is due to the mechanical stress of the insulator. These two can be difficult to distinguish, but the triboelectric effect is expected to have higher contribution.\cite{KalraArxiv2016} Furthermore, bending cables with highly resistive insulators such as teflon (PTFE) can introduce charge traps that decay as rapid spikes over a very long time ($>24$ hours). In addition to the mechanical vibrations, also temperature oscillations of the pulse tube can affect current measurements through the thermoelectric effect in the cables.

The frequency range of interest can be divided into three categories: \emph{(i)} Millihertz frequencies are most crucial for quantum current metrology, because the current reversal frequency is limited by the bandwidth and $1/f$ noise of amplifiers.\cite{DrungRevSciInstr2015,SteinAPL2015,GiblinNatCom2012} \emph{(ii)} The few Hz range is important for the characterization of quantum devices by measuring small dc electric currents. The minimum sampling time for high-resolution dc measurements is  limited by the bandwidth of the amplifier and often also by the power line frequency. However, in pulse tube cooled cryostats, much longer sampling times are needed if excess noise is generated by the pulse tube oscillations with frequency of the order of 1 Hz. In this case the measurement time can become devastatingly long. \emph{(iii)} The audio frequency range is relevant for many types of ac experiments, especially if the noise can affect the studied quantum system itself.\cite{KalraArxiv2016} In this paper, we focus on suppressing current noise in the frequency ranges \emph{(i)-(ii)}, but we also demonstrate the usefulness of those methods for audio frequency experiments.


\section{Experimental methods}

The measurements were performed in a BlueFors BF-SD250 cryogen-free dilution refrigerator. It uses a Cryomech PT410 pulse tube cryocooler to provide sufficient temperature for pre-cooling and  condensing the mixture of $^3$He and $^4$He. The pulse tube has a repetition rate of 1.4 Hz. Pulsing the helium causes mechanical vibrations from 1.4 Hz up to kHz range in the connected structures. The remote motor and the compressor of the pulse tube are outside the shielded measurement room. The pulse tube cold finger is rigidly connected to a room temperature vacuum flange of the cryostat. Inside the cryostat, the 1st and 2nd stages of the pulse tube are connected with flexible copper braids to respective flanges, which are meant for thermalization of cables, He lines, radiation shields etc. Originally, the pulse tube vibrated the whole cryostat assembly. Its support structures were substantially reinforced. After the modifications, the low frequency noise due to the vibrations of the support structure became negligible. Therefore, we concluded that the remaining vibrations are due to mechanical coupling of the pulse tube to the cryostat flanges  through the copper braid thermal contacts. We note that the diameter of the sample space of our cryostat (15~cm) is quite small, which makes the whole cold finger more prone to vibrate. Because the rigidity of a hollow cylinder increases as the 3rd power of the diameter (when the wall thickness is fixed), the mechanical vibrations of our cryostat can be even an order of magnitude higher than in larger pulse tube based dilution refrigerators.

\begin{figure}[ht]
\includegraphics[width=\columnwidth]{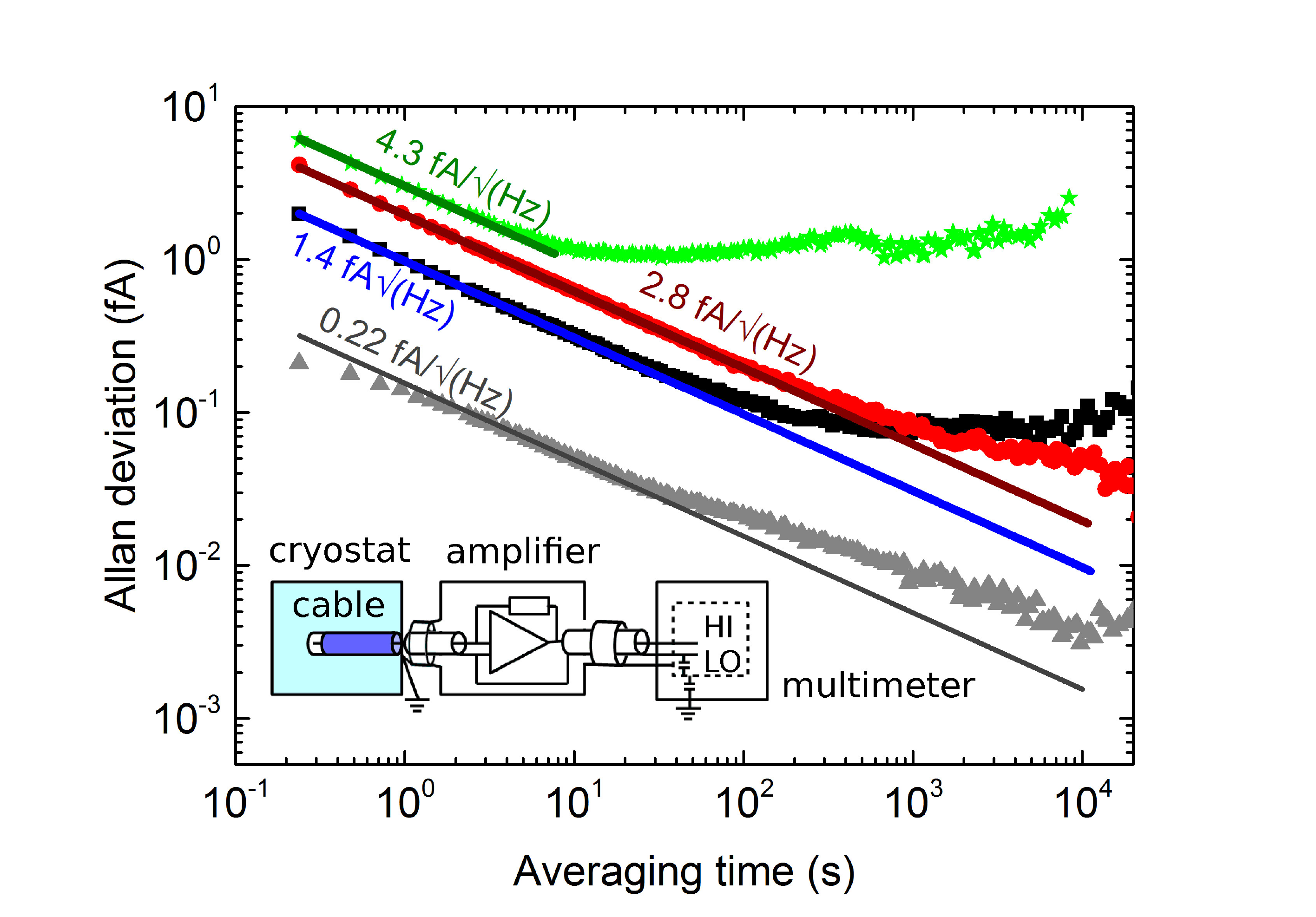}%
\caption{\label{mittarit} Allan deviations of the amplifiers ULCA (red circles), Femto DDPCA-300 at gain $10^{11}$  V/A (black squares), Femto DDPCA-300 at gain $10^{12}$ V/A (gray triangles) and Femto DLPCA-200 (green stars), and white noise fits to the beginning of data. Inset: measurement setup.}%
\end{figure}

The measurement setup can be seen in the inset of figure \ref{mittarit}. One end of the cable under test was connected to a transimpedance amplifier and the other end was a shielded open circuit. The amplified signal was digitized with an Agilent 3458A multimeter using different sampling times and time trace lengths. The vacuum flange of the cryostat was used as the grounding position of the experiments. The room temperature cabling was triaxial: coaxial cables were used as signal lines, and an additional copper braid shield was used to connect the guard shield of the multimeter to ground. This effectively suppresses the effects of capacitive ground loops. We also observed that connecting multiple amplifiers and voltage meters (for measuring multiple cables simultaneously) without computer data bus or power line isolation can introduce distortion somewhere in the circuit that corresponds up to almost 10~fA at 1.4~Hz. Adding data bus isolation also reduced noise in some of the measurements done with single multimeter. This demonstrated the importance of the electromagnetic design of sensitive current measurements.

The digitized data were later analyzed in terms of the Allan deviation,~\cite{WittEPJSP2009} which depicts the statistical uncertainty of the current measurement at certain averaging time, and frequency spectrum, which shows the noise peaks as a function of frequency. When doing long time trace measurements,  we used a sample integration time of 200 ms and performed autozero calibration of the multimeter every 5 seconds to prevent drifts of the multimeter from affecting the results.~\cite{DrungMetrologia2015} The autozero calibration caused dead time in the measurements, which was corrected in the data analysis.

We used three different transimpedance amplifiers: ULCA,~\cite{DrungRevSciInstr2015} Femto DDPCA-300 and Femto DLPCA-200. Allan deviations of the amplifiers can be seen in figure~\ref{mittarit}. In general, the noise of the transimpedance amplifier is minimized when the feedback resistor and the transimpedance gain are large like in Femto DDPCA-300. Femto DLPCA-200 is designed for higher frequencies, at the expense of higher noise. Since large resistors exhibit significant drifts in time and temperature, the Femto amplifiers are not ideal for metrological current measurements except when they are used as null instruments.~\cite{GiblinNatCom2012} The ULCA amplifier uses a two-stage amplification design which allows to use stable thin-film and metal foil resistors that can be calibrated with a cryogenic current comparator.~\cite{DrungIEEE2015} Thus, the ULCA amplifier can be used e.g.~to close the quantum metrology triangle.~\cite{ PekolaRevModPhys2013} The drawback of using smaller resistors is a lower gain (10$^9$ V/A) and somewhat higher white noise density (2.8~fA/$\sqrt{\mathrm{Hz}}$) than in Femto DDPCA-300 (1.4~fA/$\sqrt{\mathrm{Hz}}$ at gain $10^{11}$~V/A  and 0.22~fA/$\sqrt{\mathrm{Hz}}$ at gain $10^{12}$ V/A).~\cite{footnote} Therefore, Femto DDPCA-300 was the most useful device for characterizing the cables, but the noise of ULCA sets our metrological goal: the cable noise should be sufficiently low to ensure that the resulting noise of current measurements is dominated by an accurate amplifier like the ULCA. Femto DLPCA-200 had the current noise density of 4.3 fA/$\sqrt{\mathrm{Hz}}$  at its maximum gain $10^{11}$ V/A. The cut-off frequencies of Femto DLPCA-200 at gain 10$^{11}$ V/A, ULCA, and Femto DDPCA-300 at gains 10$^{11}$~V/A and 10$^{12}$~V/A are about 1~kHz, 60~Hz, 20~Hz, and 1~Hz, respectively.

One should also note that the results of figure~\ref{mittarit} were obtained in rather ideal conditions: The laboratory temperature was stabilized, and the amplifiers were battery powered using external regulators, which further suppresses temperature drifts. Even in these conditions Femto DDPCA-300 shows drifts from tens of minutes to several hours after short-circuited for sample protection and up to 24 h after switched on. The presented data was gathered when the initial drifts had become small.

\begin{figure}[ht]
\includegraphics[width=\columnwidth]{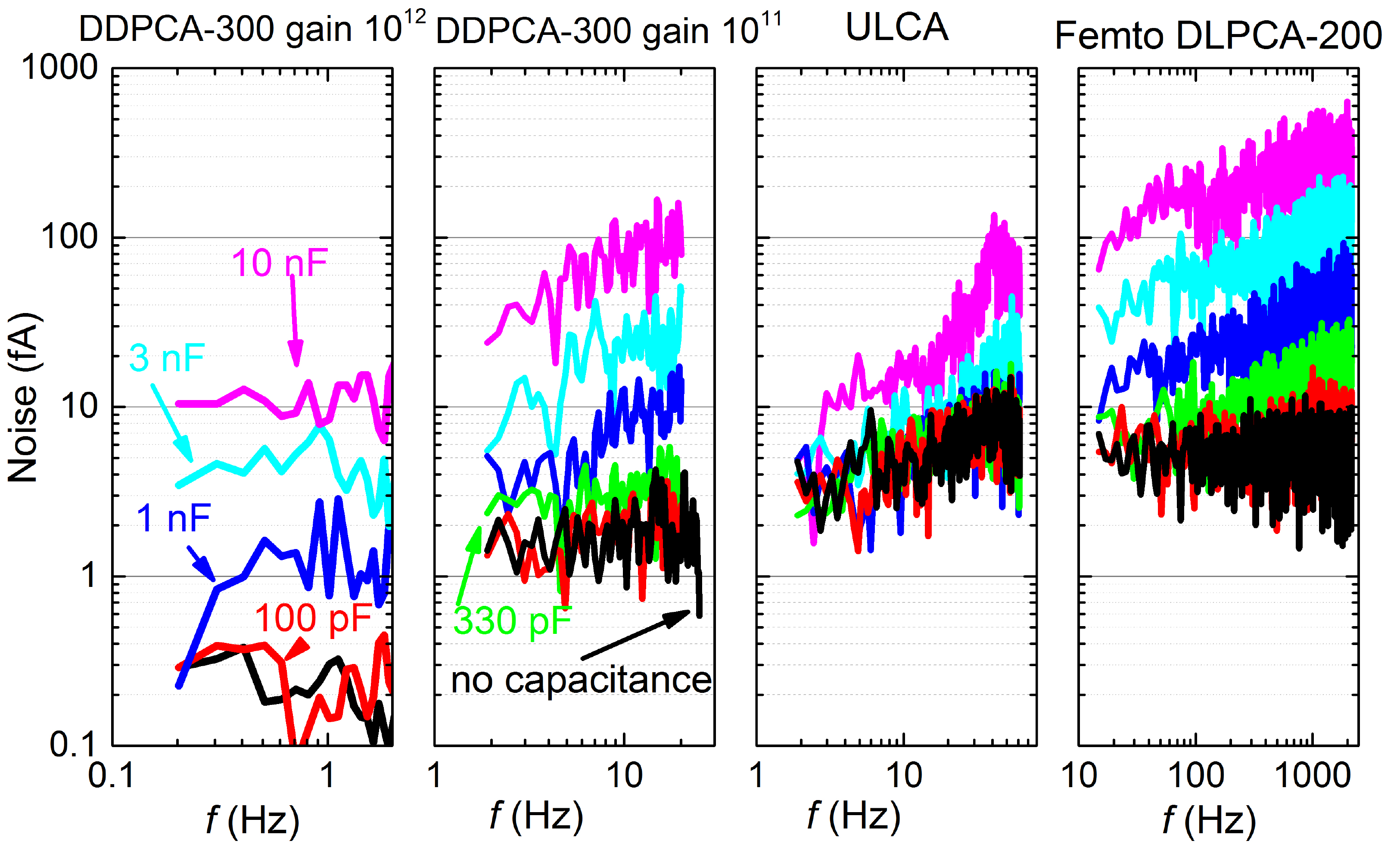}%
\caption{\label{konkat} The effect of capacitance in the input of the three amplifiers. The measurements were performed without capacitance (black line) and with capacitances 100 pF (red line), 330 pF (green line), 1 nF (blue line), 3nF (cyan line) and 10 nF (magenta line). For each subfigure the upper frequency is set by the amplifier cut-off frequency. }%
\end{figure}

In addition to the noise components described above, the cable tests are also affected by voltage noise of the amplifiers, since it will introduce additional current noise due to the cable capacitance. In cryogenic environments we need relatively long cables and can therefore have also noticeable capacitances. The effect of a capacitor connected to the input of the amplifiers is presented in figure~\ref{konkat}. One can see that ULCA performs better than or as well as Femto DDPCA-300 at gain $10^{11}$ V/A when the capacitance is 1 nF or higher.  In addition, if the capacitance is 3 nF or higher, the noise of ULCA at 2 Hz is lower than or equal to the noise of Femto DDPCA-300 with gain $10^{12}$ V/A at 0.2~Hz -- 1~Hz. We have summarized the capacitances of the tested cables in Table II in Appendix A.

\section{Cables under test}

\begin{table*}[t]
\caption{\label{cables1} Cables under study, including the insulator of each cable and the cryostat flanges between which the cable was used. The respective flanges were: room temperature vacuum feedthrough (RT), first pulse tube stage (PT1, 60~K), second pulse tube stage (PT2, 4 K), still (S, 800~mK) and mixing chamber (MC, 20~mK) flanges.}
\begin{tabular}{|l|p{3cm}|p{3.5cm}|p{4.5cm}| p{4.5cm} |}
\hline
cable & insulator & cryostat flanges & pros & cons \\
\hline
CMR  & polyimide + PTFE 	& RT - MC & easily available  & noisy \\
GVL1 	& PTFE + graphite 	& RT - MC or PT2 - MC &  easily available &\\
GVL2 	& PE		 	& RT - MC & easily available & \\
SMPE	& PE + PVC		& RT - MC &  & self made, breaks easily\\
VI$_1$/ VI$_2$ & vacuum & RT -  PT2	& vacuum as insulator &  cannot be bent\\
HS$_1$/ HS$_2$	& PE + PVC		& RT 		& easily available &room temperature cable\\
TC	& magnesium oxide	&  PT2 - MC & filters high frequency noise & hygroscopic insulator\\
\hline
\end{tabular}
\end{table*}

In room temperature low noise cables, the insulator is typically either polytetrafluoroethylene (PTFE) coated with graphite or polyethylene (PE) coated with polyvinyl chloride (PVC). Though good cables exist for room temperature experiments, the situation is much worse in cryogenic environments, where standard conductors cannot be used due to their high heat conductivity. In this article, we investigate the current noise of several cable setups based on custom-made, standard commercial and self-made components to find a low-noise setup without sacrificing other cryogenic requirements. A short overview of the cables under test is presented in table \ref{cables1}. A more thorough list of properties can be found in appendix A.

Initially, we used shielded twisted pair cables from CMR-Direct (abbreviated hereafter as CMR). They had a PTFE layer under the shield, and the individual lines were coated with polyimide. Their noise was not satisfactory for our experiments with quantum current sources and thus we acquired two low-noise cables from GVL Cryoengineering. One of them (GVL1) had graphite coated PTFE and the other one (GVL2) polyethylene as insulator. We also fabricated a multicore cable by inserting Isa-Ohm\textregistered{} (see Appendix A for specifications) wires inside PE tubes (SMPE, self-made, polyethylene insulator). All these cables could be used to cover the whole distance from a room temperature vacuum feedthrough to the mixing chamber (MC) flange of the cryostat.

\begin{figure}[ht]
\includegraphics[width=\columnwidth]{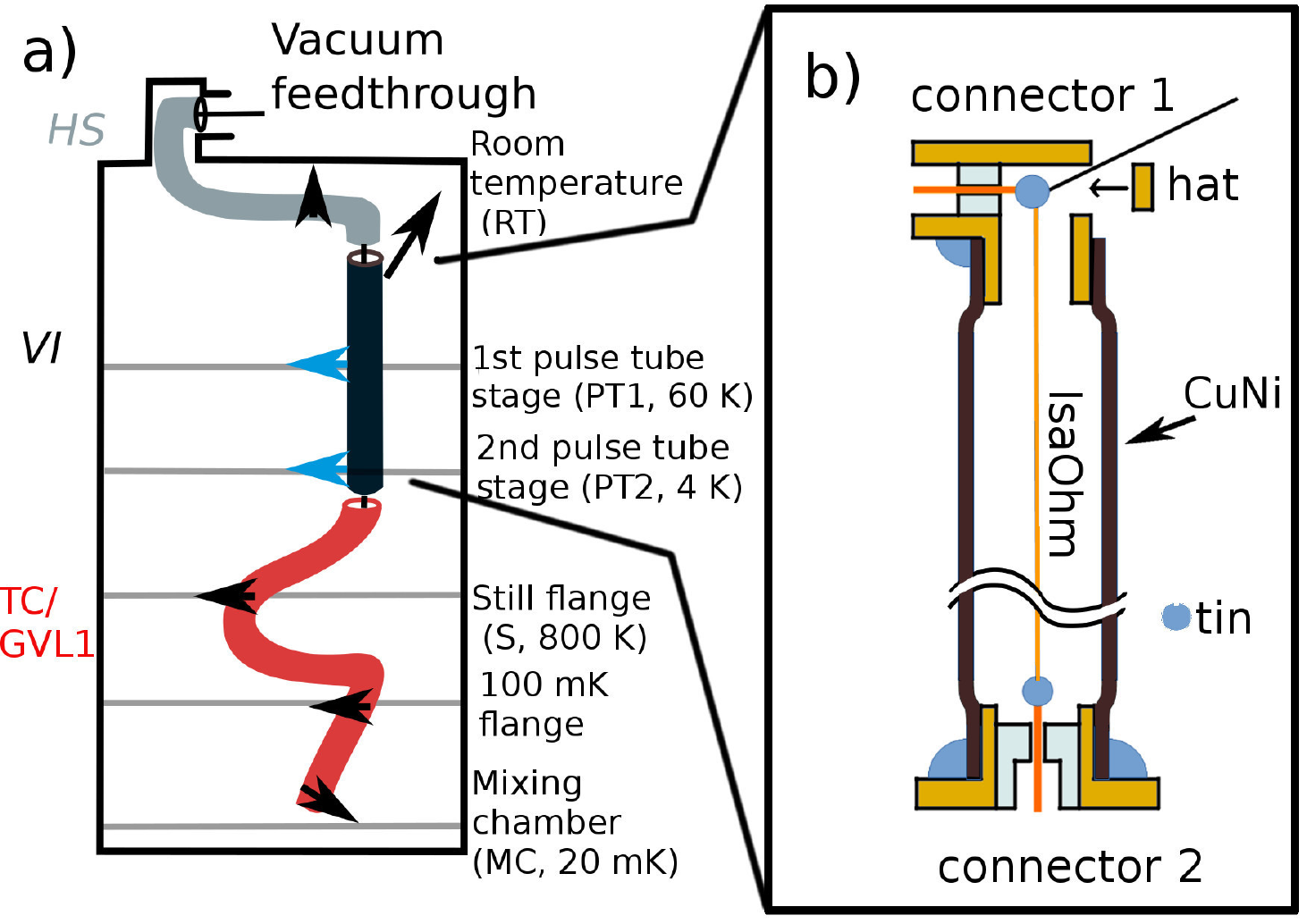}%
\caption{\label{vi} \color{mycol} a) Schematics of HS+VI+TC/GVL1 cable inside the cryostat. Arrows represent thermalizations. A room temperature cable HS (gray) was used to carry signal from vacuum feedthrough to vacuum insulated cable VI (black), since the geometry of the cryostat required flexible cables. Cable VI was used between room temperature and the 2nd pulse tube stage (PT2). Its outer conductor was thermalized to respective flanges of the cryostat. We used TC or GVL1 (red) between the PT2 and mixing chamber (MC) flange.  b) Shematics of the VI cables (not to scale). Connector 1 had an opening through which the inner conductor could be tightened. The conductor was later cut to fit inside the connector and the opening was shielded with a hat. The connectors were connected to the CuNi tube by first crimping and then soldering.}
\end{figure}

Since the noise properties of GVL and SMPE cables were not completely satisfactory, we made two vacuum insulated coaxial cables (VI$_1$ and VI$_2$), see figure \ref{vi}b. These cables feature the same general design but differ in outer conductor diameter (4 mm for VI$_1$ and 8 mm for VI$_2$). They cannot be bent and therefore we used a low noise room temperature cable from Huber+Suhner (HS) between the cryostat vacuum feedthrough and the VI cables. The HS cable was inside the cryostat but completely at room temperature, see figure \ref{vi}a. The center conductor of the VI cable cannot be thermalized to intermediate temperatures and cannot thus be used to carry signal directly from room temperature to the MC flange. One could use multiple vacuum insulated cables and thermalize the center wire between them, but we used either GVL1 or Thermocoax\textregistered{} (TC) between 2nd pulse tube stage (PT2) and the MC flanges.

We had observed in room temperature measurements that TC is relatively insensitive to vibration induced noise. However, it can exhibit thermoelectric voltages of the order of a millivolt between room and cryogenic temperatures. These voltages could break sensitive samples if they are not compensated by using identical lines in both sides of the sample. In addition, MgO, the insulator of TC, is highly hygroscopic and becomes weakly conducting at room temperature if it is exposed to humidity. To avoid these problems, we used TC only between the PT2 and MC flanges.


\section{Experimental results}

Below we study three different frequency ranges introduced earlier. The millihertz range is studied with Allan deviation calculations.~\cite{WittEPJSP2009} This range is most crucial for precise quantum metrological measurements, where long averaging times are needed. The dominant noise sources are intrinsic noise of the cables due to coupling to the environment etc., the intrinsic noise of the amplifier, flicker noise and in some cases also pulse tube noise. The few hertz region is studied with Fourier analysis and is important for fast characterization of samples. The main noise source are the pulse tube vibrations. For VI cables the Fourier analysis is extended up to 50~kHz, where significant pulse tube induced noise has been observed in other experiments.~\cite{KalraArxiv2016}

\subsection{Allan deviations}

\begin{figure}[ht]
\includegraphics[width=0.9\columnwidth]{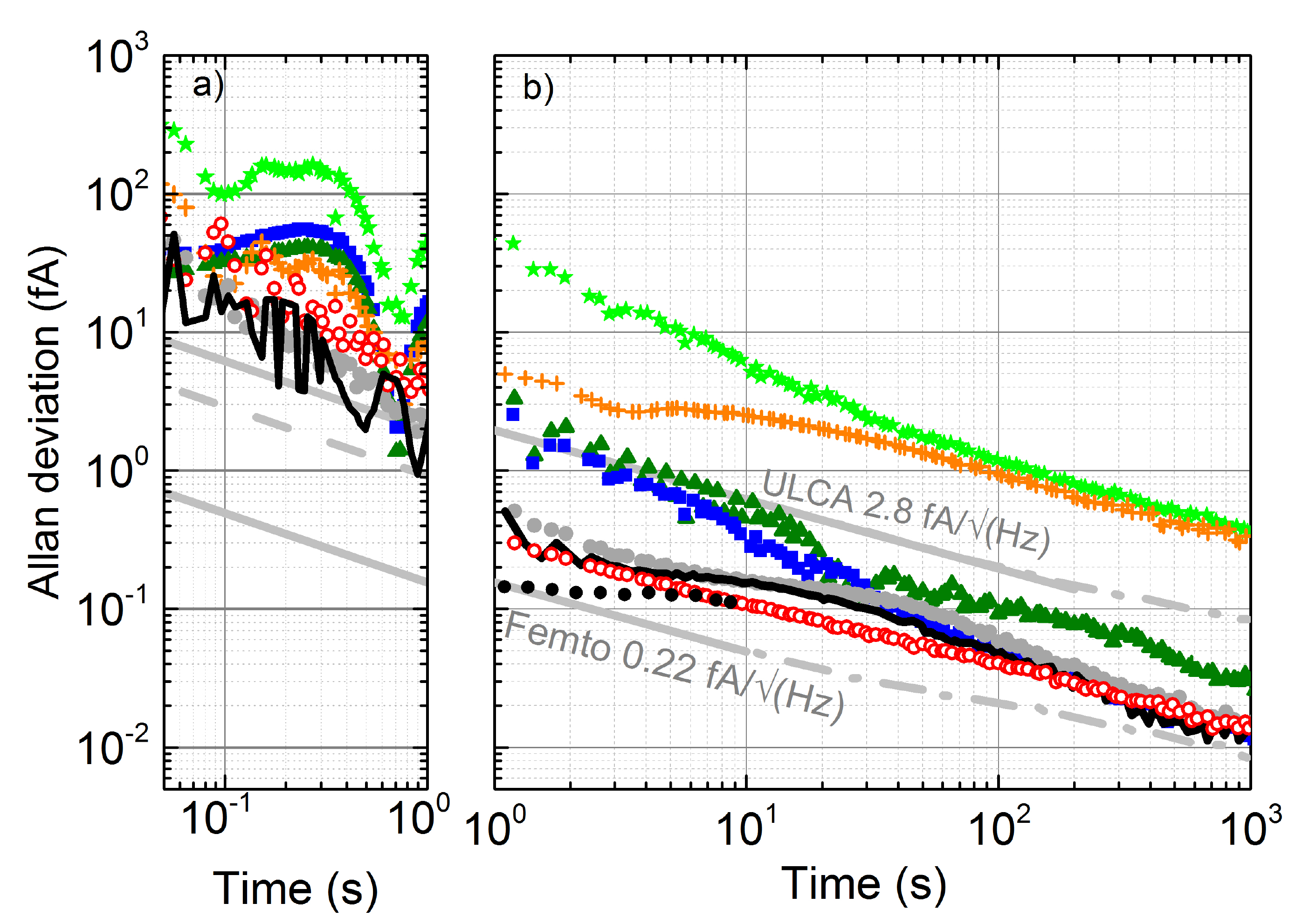}%
\caption{\label{allan} Allan deviations of noise currents for full length wires at cryogenic temperatures and under the influence of pulse tube. Measurements were performed using Femto DDPCA-300 gains a) $10^{11}$ V/A and b) $10^{12}$ V/A and with ULCA (light green stars and orange crosses in b) . The measured cables were HS$_1$+VI$_1$ (black solid line), HS$_1$+VI$_1$+TC (gray solid circles), HS$_2$+VI$_2$+GVL1 (red open circles), SMPE (blue squares), GVL1 (dark green triangles), GVL2 (orange crosses) and CMR (light green stars). The solid gray lines present the white noise of Femto DDPCA-300 at gain $10^{12}$ V/A and that of ULCA and the dashed gray line that of Femto DDPCA-300 at gain $10^{11}$ V/A. The dash dotted lines present the measured noise of  Femto DDPCA-300 at gain $10^{12}$ V/A and that of ULCA in the region where the noise is not white. The data described with black dots in b) was measured when the pulse tube was off and represents cable HS$_1$+VI$_1$.}%
\end{figure}

Figure \ref{allan} presents the Allan deviations for different cables that extend from the room temperature vacuum feedthrough down to the MC flange of the cryostat, or to the PT2 flange in case of HS+VI, see table \ref{cables1} for details. The measurements were made using the ULCA and the Femto DDPCA-300 at gains 10$^{11}$ V/A and 10$^{12}$ V/A. Possible reasons for the discrepancy at 1~Hz between figures \ref{allan}a and \ref{allan}b are different voltage noise contributions of the Femto DDPCA-300 at different gains, and also the 1~Hz cut-off frequency with the larger gain. The drop around 0.7~s in figure \ref{allan} originates from the fact that pulse tube noise is supressed when the averaging time corresponds to its noise cycle.

 Figure \ref{allan}b demonstrates that the lowest noise is obtained when the VI cable is used between the room temperature and the PT2 flange at 4~K. Since a single VI cable could not be used to carry signal directly from room temperature to base temperature of our cryostat (MC flange), we extended it either with GVL1 or TC between PT2 and MC flanges, see figure \ref{vi}a. Interestingly, adding GVL1 or TC after VI cables does not affect the Allan deviation above 1 s averaging times (black line, red open circles and gray dots in figure \ref{allan}), even though GVL1 has significant amount of noise when used between room temperature and MC flange (dark green triangles in figure \ref{allan}). From a fit to the first 400 s of Allan deviation of HS$_2$+VI$_2$+GVL1, we obtain white noise density of 0.44 fA/$\sqrt{\mathrm{Hz}}$. Corrections due to dead times are made in the results. Also, using the GVL1 and SMPE cables leads to noise which is lower than the intrinsic noise of the ULCA if the averaging time is above 10 s.

We also tested the effect of the pulse tube to the Allan deviations. The black dots in figure \ref{allan} present a measurement of HS$_1$+VI$_1$, when the pulse tube was switched off. The data were acquired at the beginning of a warm up sequence. The measured current data drifted significantly, presumably due to the rapidly increasing temperature. Therefore, only a short time trace could be used and already at 10~s, the Allan deviation is affected by the drift. However, we notice that in the range between 1 s and 10 s the Allan deviation of HS$_1$+VI$_1$ is lower when the pulse tube is off than when it is on, indicating that some amount of pulse tube noise penetrates down to the subhertz range.

\subsection{Frequency domain}

\begin{figure}[ht]
\includegraphics[width=\columnwidth]{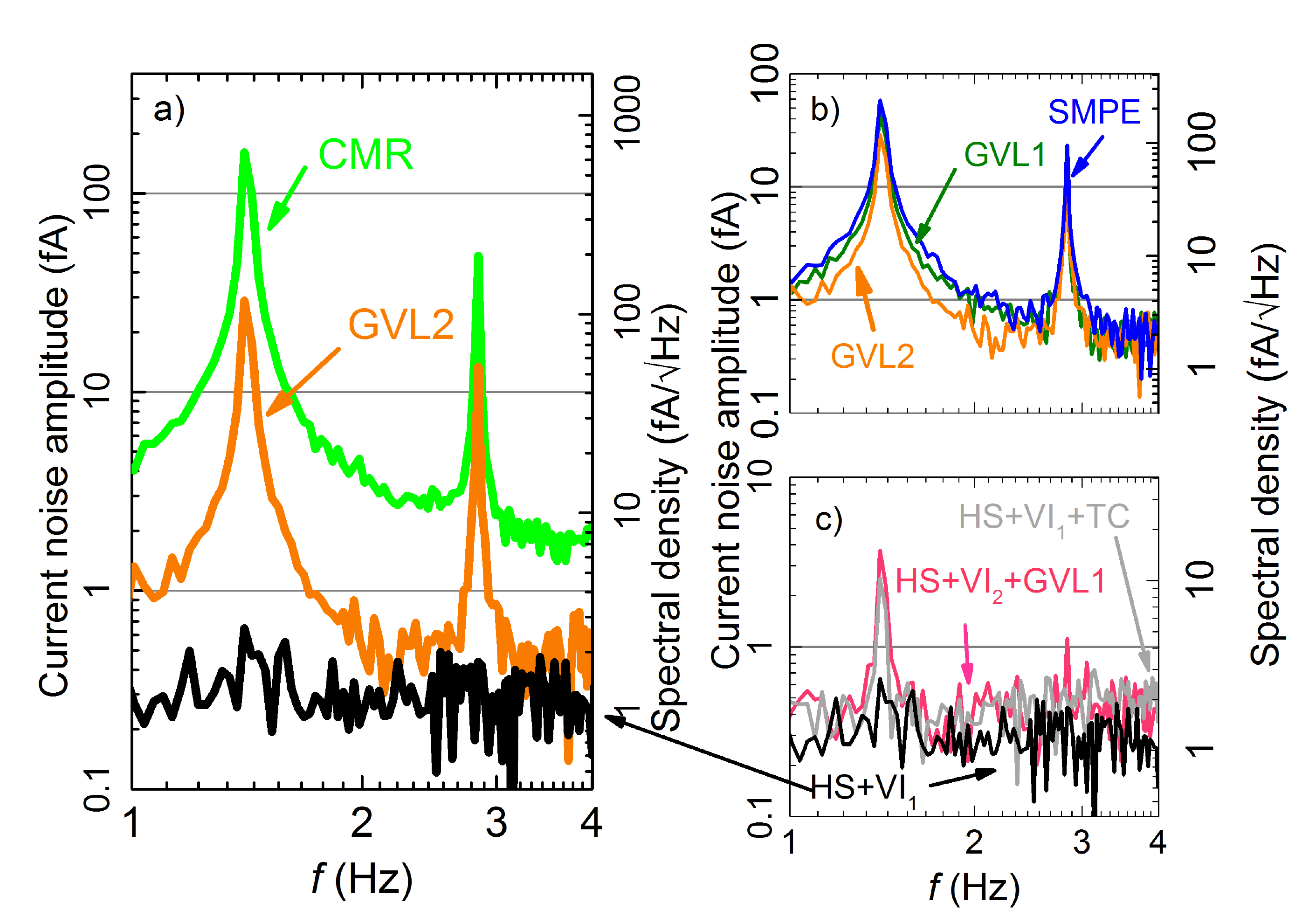}%
\caption{\label{fft1} Frequency spectra between 1 Hz and 4 Hz showing the 1.4 Hz peak originating from pulse tube vibrations. The cables were measured at cryogenic temperatures and under influence of pulse tube. Left axis: scaling for peak amplitudes. Right axis: scaling for background amplitude spectral density. a) Comparison between HS$_1$+VI$_1$, GVL2 and CMR. b) Comparison between SMPE, GVL1 and GVL2. c) Comparison between low noise cables: HS$_1$+VI$_1$, HS$_1$+VI$_1$+TC and HS$_2$+VI$_2$+GVL1.}
\end{figure}

\begin{figure}[ht]
\includegraphics[width=\columnwidth]{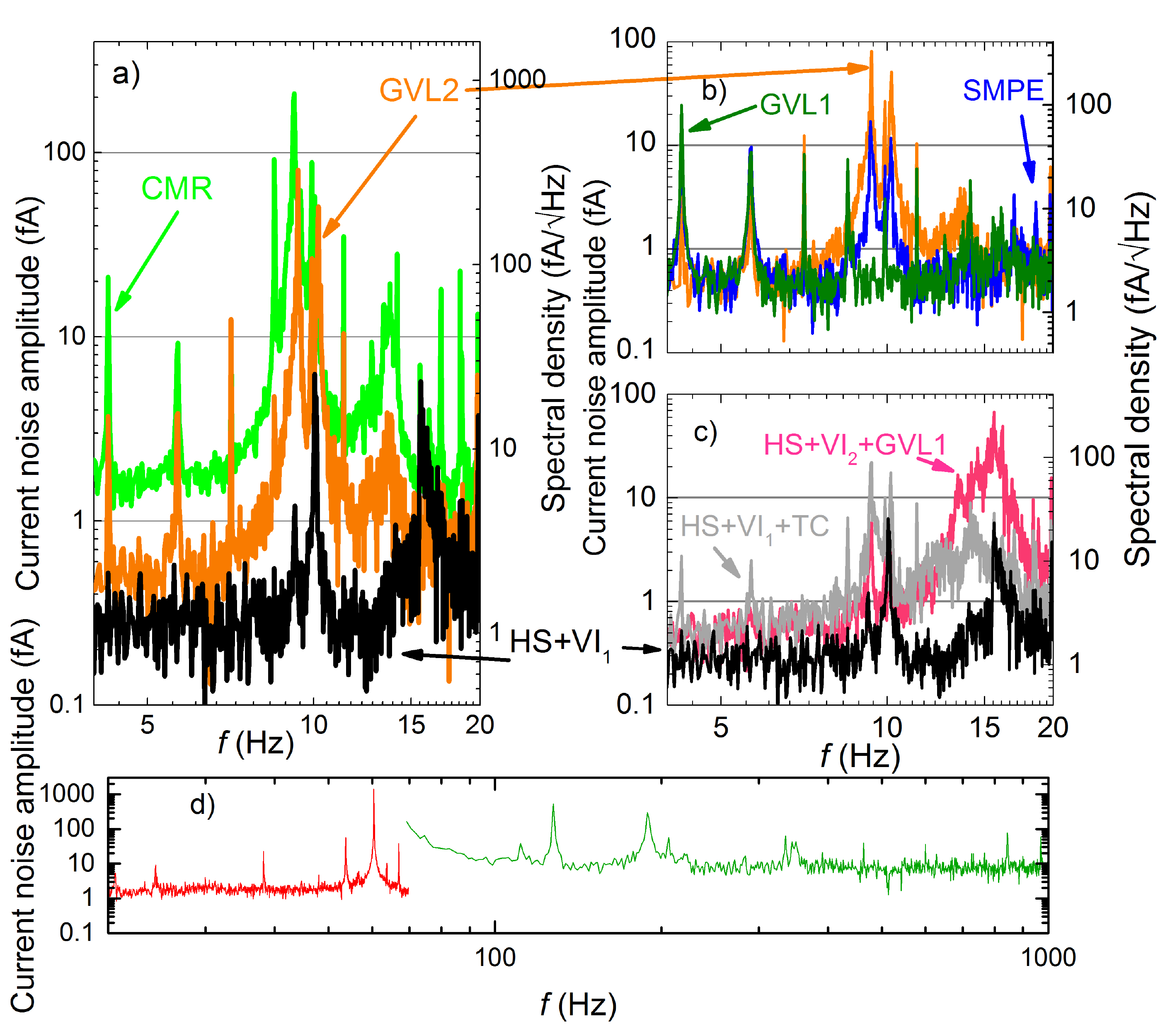}%
\caption{\label{fft2} Frequency spectra of the cables at cryogenic temperatures and under influence of pulse tube between 4 and 20~Hz. a) Comparison between HS$_1$+VI$_1$, GVL2 and CMR. b) Comparison between SMPE, GVL1 and GVL2. c) Comparison between low noise cables: HS$_1$+VI$_1$, HS$_1$+VI$_1$+TC and HS$_2$+VI$_2$+GVL1 d) Frequency spectrum from 20 Hz to 1 kHz for HS$_1$+VI$_1$ measured with ULCA (red line, from 20~Hz to 70~Hz) and Femto DLPCA-200 (green line, from 70~Hz to 1~kHz ).}
\end{figure}

Frequency spectra of the current noise of the cables can be seen in figures \ref{fft1} and \ref{fft2}, where scaling for both peak amplitude and frequency spectral density are shown.~\cite{NIfft} 

The dominant features in figure \ref{fft1} are the pulse tube generated peaks at 1.4 Hz and 2.8 Hz. These peaks are not present in  HS$_1$+VI$_1$. In addition,  HS$_1$+VI$_1$+TC and HS$_2$+VI$_2$+GVL1 have relatively low 1.4~Hz peaks of about 4 fA and almost negligible 2.8 Hz peaks. This is an essential improvement compared to the other cables which have 1.4~Hz peak amplitudes from about 20~fA up to more than 100~fA. In addition, HS$_2$+VI$_2$+GVL1 has much lower 1.4 Hz and 2.8 Hz peaks than bare GVL1, indicating that even though a cable would exhibit high noise current at pulse tube frequency  when extended the full length from room temperature to the MC flange (as is with bare GVL1)  the same cable might work well between PT2 and MC flanges (as is with GVL1 in the combination cable HS$_2$+VI$_2$+GVL1, see figure \ref{vi}a). It should be noted, however, that we saw an increase in 1.4 Hz and 2.8 Hz peak height and even a peak when measuring  HS$_1$+VI$_1$, when the experimental setup was not done with care. The likely source of the noise were BNC connectors used in the measurement. Either the contact between the outer shells or the piezo- and triboelectric effects of connectors may give rise to this increase. 

Figure \ref{fft2} shows that HS$_1$+VI$_1$ works relatively well almost up to 50 Hz, but when combined with GVL1 or TC, the results are not as good. However these cables exhibit low noise densities up to almost 10~Hz, which allows fast enough high resolution dc-characterization for typical experiments. We notice several peaks in the spectrum of  HS$_1$+VI$_1$, first non-multiple of 1.4 Hz being at 10 Hz and 15 Hz and the ones with highest amplitude being at 60 Hz and around 130 Hz and 190 Hz. The peaks at 10 Hz and 15 Hz are likely due to mechanical vibrations of the support structure.  Most of the peaks above 15~Hz are not changed when the pulse tube is switched off, however the 60~Hz peak becomes smaller. This indicates that the origin of these peaks is mostly not pulse tube induced. We also measured the frequency spectrum of HS$_1$+VI$_1$ up to 50 kHz (data not shown). Between 1 kHz and 50 kHz the noise was dominated by that of the Femto DLPCA-200 amplifier (about 100~fA$/\sqrt{\mathrm{Hz}}$) and no peaks were visible.
Even as there is significant amount of current noise at higher frequencies, it is still smaller than e.g.~in the experiments of~\textcite{KalraArxiv2016}  Therefore, even though our setup was aimed for lower frequencies, the vacuum insulated cable technology is also promising e.g. for qubit experiments at higher frequencies.

We also investigated the contribution of thermoelectric effect on the pulse-tube-generated current noise in HS$_1$+VI$_1$ but found it to be negligible. However, we cannot completely rule out this effect especially when using TC, which has a high Seebeck coefficient.

\section{Conclusion}

We have demostrated that vacuum insulated cables are almost immune to pulse tube based vibrations. It seems to be sufficient that they are used to carry signal between the room temperature and the 2nd stage of the pulse tube. Two commercial cables yield excellent results when they are used between the 2nd stage and mixing chamber temperatures. The performance of the cabling solution is expected to be even better in modern pulse tube based dilution refrigerators with larger diameter than ours.

In the millihertz range, the Allan deviation analysis shows that our lowest cable noise equals roughly to the white noise density of
0.44~fA$/\sqrt{\mathrm{Hz}}$, which is much smaller than that of the ULCA amplifier (2.4~fA$/\sqrt{\mathrm{Hz}}$).
Therefore, cables do not limit the uncertainty of a quantum metrology triangle experiment performed with ULCA.\cite{DrungIEEE2015}

We studied the few hertz range in frequency domain using Fourier analysis. In vacuum insulated cable, the noise amplitude of pulse tube induced current distortion became vanishingly small compared to the noise floor set by the amplifiers. In addition, the amplitude is below 4~fA in two low-noise commercial cables when they are used in combination with the vacuum insulated cable. The observed noise amplitude at pulse tube frequency is smaller by about a factor of 50 compared to the original cable we used. This enables fast characterization of quantum devices when measuring small electric currents, since the sampling frequency is not anymore limited by that of the pulse tube.

Although it is not the main focus of our work, we tested the vacuum insulated cable up to 50~kHz. They have a few notable distortion peaks between 50 and 200~Hz, but above 1~kHz only the noise floor of the amplifier, 100~fA$/\sqrt{\mathrm{Hz}}$, can be seen. This result is promising for extending the use of the VI cabling solution for higher frequency applications, for example spin quantum bit experiments.

\begin{acknowledgements}

We thank Chao Wang and Jari H\"{a}llstr\"{o}m for useful discussions and George V. Lecomte for providing custom-made cables. This research was financially supported by the Wihuri foundation (E.M.) and the grants no.~288907 (J.S.L. and A.J.M.)  and 259030 (A.K.) of the Academy of Finland.
\end{acknowledgements}



%


\

\section{Appendix A}
Table \ref{cables} depicts the investigated cables, including the lengths, resistances per unit length (res.), diameters of the inner and outer conductors (dia.) and capacitances per unit length (cap.) of the cables.

\begin{widetext}
\begingroup
\begin{table*}[ht]
\caption{\label{cables} Appendix: Investigated cables.}
\begin{tabular}{|p{1.0cm} | p{1.1cm} | p{1.7cm} | c | p {1.2cm}| p{1.7cm}  | c | c | p{2.7cm}| c | p{2.7cm}| }
\hline
	&  & \multicolumn{3}{|c|}{\textbf{inner conductor}} &\multicolumn{3}{|c|}{\textbf{outer conductor}} & \multicolumn{2}{|c|}{\textbf{insulator}}& \textbf{product number} \\
\cline{3-10}
\textbf{cable} &\textbf{length} 	& \textbf{material} & \textbf{res.} & \textbf{dia.}  & \textbf{material} & \textbf{res.} & \textbf{dia.}& \textbf{material}  &\textbf{cap.} & \\
		 	 & 	    cm &				&$\Omega/$m	  &mm	&			& $\Omega/$m 	&mm			&				& pF/m &\\
\hline
CMR		&	200 & brass 				&	8.1	& unknown	& CuNi braid	&	5.5		&	0.8		&polyimide~+~PTFE	&  	70 		&MSC-2BR \\
GVL1 		& 200&	 CuNi				&	92	&	0.07		&	CuNi		&	2.8		&	0.6		&PTFE~+~graphite		&	90		&	GVLZ185	\\
GVL2 		& 300 &	brass				&	 4.4	&$2\times 0.1$	& CuNi braid	&	5.5		&	0.8		&	 PE			&	 84 		&	GVLZ189	\\
SMPE 		& 200 &Isa-ohm\texttrademark 	& 134	&0.112		& \textendash 	&\textendash	&\textendash	&PE~+~PVC 			&  varies		&\textendash\  	\\
VI$_1$	& 20	&Isa-ohm\texttrademark	& 134 	&0.112		& CuNi		&	2.5		&	4		&	vacuum		& 	unknown	&\textendash 	\\
VI$_2$	& 30	&Isa-ohm\texttrademark 	& 134 	&0.112		& CuNi		&	1		&	8		&	vacuum		& 	unknown	&\textendash 	\\
HS$_1$  		& 45	&	Cu				&unknown	&	0.489	&	Cu		& 	unknown	&	4		&PE~+~PVC			&	71		&  Huber$+$Suhner\\	
HS$_2$  		& 65	&					&	&		&			& 		&			&			&			&   G\_03130\_HT-01\\	
TC 		& 100	&	NiCr				&	50	&0.17		&	stainless steel	&	6.9		&	0.5		&magnesium oxide	&	490		& thermocoax\textregistered \  \ 1Nc Ac 0.5\\
\hline
\end{tabular}
\end{table*}
\endgroup
\end{widetext}

\bibliography{reducingNoiseInCryoExperiments}

\begin{thebibliography}{16}
\bibitem [{foo()}]{footnote}%
  \BibitemOpen
  \href@noop {} {}\bibinfo {note} {The noise floor of ULCA is 2.4
  fA/$\sqrt{\mathrm{Hz}}$. The excess noise of our measurement is most probably
  due to the multimeter. Higher gain can be obtained by using an external
  feedback resistor at the 2nd amplification stage of ULCA, but care must be
  taken not to deteriorate the absolute uncertainty of the
  measurement.}\BibitemShut {Stop}%
\end{thebibliography}

\end{document}